\documentclass{emulateapj}

\usepackage{graphicx}

\slugcomment{}

\shorttitle{The host galaxies of type-2 quasars}
\shortauthors{Lacy et al.}

\begin{document}


\title{Large amounts of optically-obscured star formation in the host
galaxies of some type-2 quasars.}


\author{M.\ Lacy\altaffilmark{1} 
A.\ Sajina\altaffilmark{1},  
A.O.\ Petric\altaffilmark{1,2},
N. Seymour\altaffilmark{1},
G.\ Canalizo\altaffilmark{3}, 
S.E.\ Ridgway\altaffilmark{4}
L.\ Armus\altaffilmark{1}, 
L.J.\ Storrie-Lombardi \altaffilmark{1}}
\altaffiltext{1}{Spitzer Science Center, Caltech, Mail Code 220-6,
Pasadena, CA 91125; mlacy@ipac.caltech.edu,sajina@ipac.caltech.edu, 
lisa@ipac.caltech.edu}
\altaffiltext{2}{Columbia University of New York, NY, USA; andreea@ipac.caltech.edu}
\altaffiltext{3}{University of California, Riverside, CA, USA; gabriela.canalizo@ucr.edu}
\altaffiltext{4}{NOAO, Colina El Pino s/n, Casialla 603, La Serena, Chile; 
seridgway@ctio.noao.edu}

\begin{abstract}
We present {\em Hubble Space Telescope} images, and spectral energy
distributions from optical to infrared wavelengths for a sample of 
six $0.3<z<0.8$ 
type-2 quasars selected in the mid-infrared using data from the 
{\em Spitzer Space Telescope}. All the host galaxies show some signs of 
disturbance. Most seem to possess dusty, star-forming disks. The disk 
inclination, estimated from the axial ratio of the hosts,  
correlates with the depth of the silicate feature in the mid-infrared
spectra, implying that at least some of the reddening towards the AGN 
arises in the 
host galaxy. The star formation rates in these objects, as inferred
from the strengths of the PAH features and far-infrared continuum, range
from $3-90 M_{\odot}{\rm yr^{-1}}$, but 
are mostly much larger than those inferred from the [O{\sc ii}]3727 
emission line luminosity, due to obscuration. 
Taken together with studies of type-2 quasar hosts from samples
selected in the optical and X-ray, this is consistent with 
previous suggestions that two types 
of extinction processes operate within the type-2 quasar population, namely
a component due to the dusty torus in the immediate environment of the 
AGN, and a more extended component due to a dusty, star forming disk. 

\end{abstract}



\section{Introduction}

Co-evolution of black holes and their host galaxies is a prediction
of most models for the observed close correlation between black hole mass
and bulge luminosity/velocity dispersion in local galaxies (e.g.\ Kauffmann
\& Haehnelt 2000; di Matteo, Springel \& Herquist 2005). 
Studies of local AGN and type-1 (unobscured) quasars show a correlation 
between AGN luminosity in the mid-infrared and starburst luminosity, estimated
from the far-infrared emission and/or PAH features
(Schweitzer et al.\ 2006; Shi et al.\ 2007; Maiolino et al.\
2007), and the host galaxies of quasars with strong far-infrared excesses
do seem to be forming stars (Canalizo \& Stockton 2001). 
However, in general, host galaxy imaging studies in the optical 
and near-IR do not suggest that
strong star formation occurs simultaneously with quasar activity in 
the majority of quasars. Imaging of type-1 quasar host galaxies show that 
the hosts of luminous quasars tend to be elliptical 
galaxies with little obvious signs of disturbance in {\em Hubble
Space Telescope} (HST) images (e.g.\ Floyd et al.\ 2004), and only about
30\% of quasars show signs of mergers or interactions in HST or adaptive
optics images (Marble et al.\ 2003; Guyon et al.\ 2007), 
though deeper (several orbit) {\em HST}/ACS imaging often reveals
faint tidal features in quasar hosts (Canalizo et al.\ 2007; 
Bennert et al., in prepraration). Studies of type-2 (obscured) 
quasars have so far been
consistent with this picture, and with the idea that the obscuration
takes place in a nuclear torus. Sturm et al.\ (2006), in 
their {\em Spitzer} study of X-ray selected type-2 quasars, make the 
claim that ``type-2 quasars are not ULIRGs'' based on featureless 
mid-infrared spectra. {\em HST} imaging studies
of the host galaxies of type-2 quasars selected from the 
Sloan Digital Sky Survey (SDSS) show that
they are consistent with being hosted by relatively undisturbed
elliptical galaxies, once the likely effects of scattered light are accounted
for (Zakamska et al.\ 2006). 

In this Letter, we present results of
{\em Spitzer} spectroscopy and photometry, and 
{\em Hubble Space Telescope} ($HST$) imaging of a sample of type-2 quasars,
and discuss the implications for the co-evolution of galaxies and black holes.
Our sample is selected in the mid-infrared,  which is less sensitive
to reddening in the host galaxy than X-ray or optical selection, and their 
type-2 nature allows us to image the central regions of the host with 
high fidelity. The combination of infrared SEDs and optical imaging 
of this sample thus allows
us for the first time 
to present a coherent picture of the nature of the star formation in 
the host galaxies of type-2 quasars.

\section{observations}

We selected all six candidate type-2 quasars with $0.3<z<0.8$ from the 
1mJy 8$\mu$m flux density limited sample
of Lacy et al.\ (2004). The redshift range ensures that these objects are
luminous enough to be on or above the quasar/Seyfert divide in the 
mid-infrared (5$\mu$m luminosity, $L_{5\rm {\mu m}} \sim 10^{23.6}{\rm 
WHz^{-1}}$, 
Lacy et al.\ 2005a), but low enough for rest-frame $B$-band to be within
the F814W filter of the Advanced Camera for Surveys
(ACS) aboard $HST$. The objects were observed with the
Multiband Imaging Photometer (MIPS) and Infrared Spectrograph (IRS) 
as part of {\em 
Spitzer} program 20083, and with HST as program GO-10848 (Table 1). 
Redshifts were obtained, and optical spectral classifications
attempted (Lacy et al.\ 2007). Low dispersion spectra were also
obtained with the Spex instrument (Rayner et al.\ 2003) 
in prism mode at the Infrared
Telescope Facility (IRTF).

\begin{table}
\caption{Observations}
{\scriptsize
\begin{tabular}{lccc}
\hline
Name & IRS & MIPS    &ACS filters\\
     & reqkey& reqkey&\\\hline
SSTXFLS J172123.1+601214&14016768&14018304&F625W,F435W\\
SSTXFLS J171147.4+585839&14016000&14017536&F814W,F555W\\
SSTXFLS J171324.2+585549&14016256&14017792&F775W,F475W\\
SSTXFLS J171831.5+595317&14016512&14018048&F775W,F475W\\
SSTXFLS J171106.8+590436&14015744&14017280&F775W,F475W\\
SSTXFLS J172458.3+591545&14017024&14018560&F775W,F475W\\\hline
\end{tabular}
}
\end{table}

The IRS data were obtained in 
stare mode, using the short-low and long-low modules. The IRS
short-low data had a background image subtracted, made by taking a median of
all the short-low observations in the Astronomical 
Observation Request. The long-low data were combined by subtracting 
nodded pairs, then combining the pairs with minmax rejection.
Extraction was performed using the SPICE tool\footnote{http://ssc.spitzer.caltech.edu/postbcd/spice.html},
with optimal extraction. Aperture photometry was performed
on the pipeline (post-BCD) MIPS mosaics using standard aperture 
corrections\footnote{http://ssc.spitzer.caltech.edu/mips/apercorr/}.

The ACS data were taken in two filters, one longward and one shortward
of the 4000\AA$\;$break in the rest frame of the galaxy. Observations
consisted of four exposures in each filter 
in a standard dither pattern, for one
orbit per object per filter. 
The ACS pipeline mosaics were also adequate for the analysis
performed here. A full analysis of the data will be presented in 
Seymour et al.\ (2007, in preparation). 

Four of our six objects were easily classified as type-2 quasars in 
Lacy et al.\ (2007) based on 
well established optical emission line diagnostics. 
However, SSTXLFS~J171106.8+590436 and SSTXFLS~J172458.3+591525 
had optical emission lines properties which were ambiguous.
For SSTXFLS172458.3+591525, we are able 
to estimate a ([S{\sc ii}]6719+6732)/(H$\alpha$+[N{\sc ii}])
ratio of 0.4 from the IRTF/Spex spectrum, which, together
with an [O{\sc iii}]/H$\beta$ ratio of 10 and assuming typical 
[N{\sc ii}]6584/H$\alpha$ ratios of 0.1-1.6, places
this object firmly amongst the AGN in the diagnostic plot 
of Kewley et al.\  (2006), with a range in 
log([S{\sc ii}]/H$\alpha$) $\approx -0.4-0.1$. 
In the case of SSTXFLS~171106.8+590436, 
the H$\alpha$ and 
[S{\sc ii}] lines are too weak to do this. However, the very high 
[O{\sc iii}]/H$\beta$ ratio of $>10:1$, detection of He{\sc ii}4686, 
and a marginal detection of [Ne{\sc v}]14.3$\mu$m in the 
IRS spectrum make it very likely that an AGN classification for the
optical spectrum is correct. We thus assume for the rest of this paper 
that all six objects contain powerful AGN.

\section{Analysis}

We fit our SEDs and spectra using a model based on those of Sajina 
et al.\ (2005, 2006). The IRS spectra were fit directly, along with 
photometric points in $r$, $i$ and $z$ bands
from the SDSS, IRAC
3.6 and 4.5$\mu$m data (Lacy et al.\ 2005b), MIPS 70 and 160$\mu$m data
(Frayer et al.\ 2006 and this paper) and near-infrared photometry from 
IRTF. We also checked the IRS
spectrophotometry against the IRAC 5.8 and 8.0$\mu$m and MIPS 24$\mu$m
(Fadda et al.\ 2006) broad-band photometry, and found them to agree within 
$\approx 10$\%.

The far-infrared (FIR) emission was fit as a modified
black body with temperature $\approx 45$K. A warm 
(small grain) component was included
with a power-law index of $\approx 2$, and cutoffs at high and low frequency.
The FIR luminosity, $L_{\rm FIR}$,
was estimated by summing the warm and cold components.
The hot dust component was based on a power-law with variable spectral index. 
The long wavelength cutoff was determined by a fit of a Fermi
function to the least starforming
AGN in our sample, SSTXLFS~J172123.1+601214, 
which shows a turndown in the
hot dust emission beginning at a rest-frame wavelength $\approx 20\mu$m. 
Netzer et al.\ (2007) also suggest that the AGN-related hot dust emission
starts to decrease beyond $\approx 20\mu$m in type-1 quasars. 
The hot dust emission was exponentially cutoff at high frequencies
with a variable cutoff frequency. The unreddened luminosity of the
AGN, $L_{\rm AGN}$, was estimated by integrating this component. 
The hot dust component is 
then reddened by the Galactic Center extinction law of 
Chiar \& Tielens (2006), 
likely to be a good approximation to the extreme density environments of 
AGN (Sajina et al. 2007). 

A modified version of the starburst 
Polycyclic Aromatic Hydrocarbon (PAH) emission 
template of Sajina et al.\ (2007) was used to fit the PAH features. AGN are
known to modify the PAH spectra (Smith et al.\ 2007). In particular, for our
objects, we found a better match by boosting the 7.7$\mu$m feature in the
template by hand. This PAH template was
fitted to the SEDs of all galaxies except SSTXFLS~J172123.1+601214,
which had a small amount of silicate emission added instead. Total 
PAH luminosities, $L_{\rm PAH}$, were determined by integrating over the
fitted template. Finally, a 5Gyr old single stellar population
from Bruzual \& Charlot (2003) was fit to the optical points. The basic
properties of the fits are listed in Table 2. 
Parameters
which are not well-constrained by the data (for example
the warm dust continuum strength in objects not detected at 160$\mu$m)
are held fixed. Full
details of the fitting process will be given in a future paper
(Ridgway et al.\ 2007, in preparation).

\begin{figure*}
\plotone{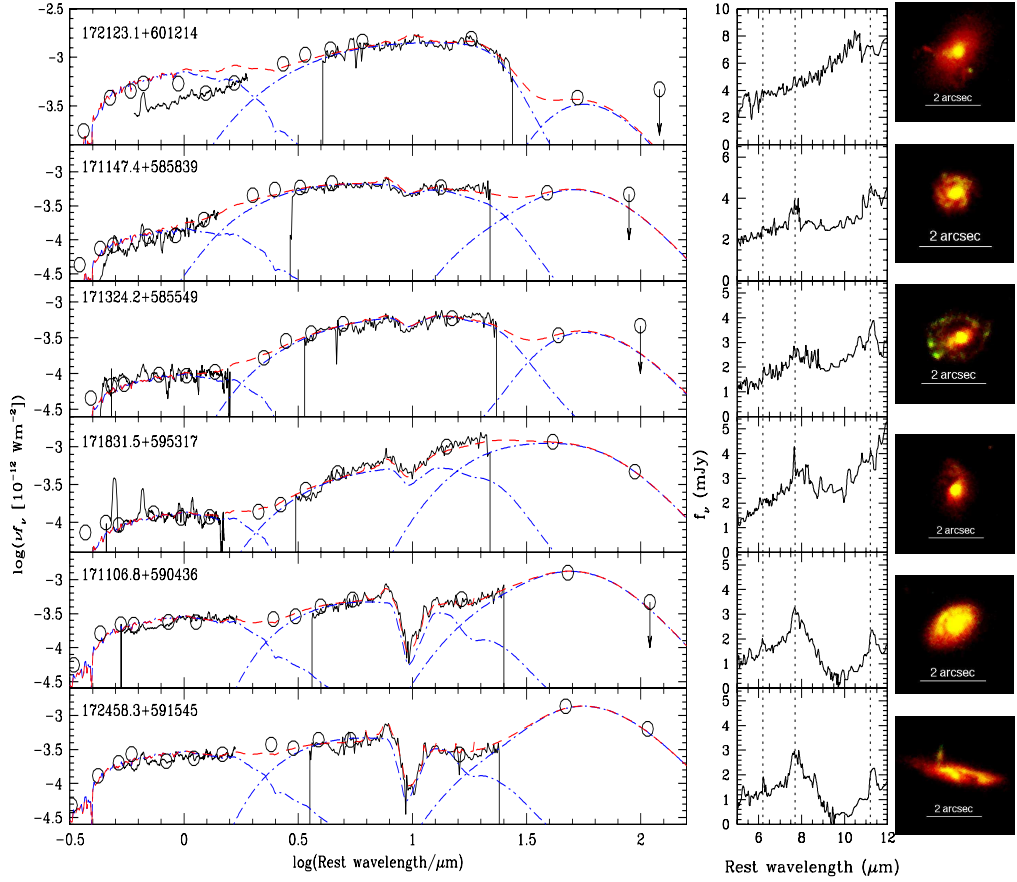}
\caption{SEDs and HST images of the six type-1 quasars, in order of increasing
silicate absorption. The solid black lines are the IRTF/Spex and 
{\em Spitzer}/IRS spectra (both smoothed to reduce noise) 
and the black open circles photometry from SDSS, IRTF and 
{\em Spitzer}. The model fit is shown as the red dashed line, with, 
from left to right, the stellar
component, hot dust component (including absorption, but excluding PAH
emission) and sum of warm and cold dust components shown
as blue chain-dotted lines. To the right of the SEDs we show expanded
plots of the IRS spectra around the PAH wavelengths. 
The ACS images have the long wavelength
image colored red and the short wavelength image in green.} 
\end{figure*}

\section{discussion}

We first address the amount of star 
formation in the host galaxies. Star formation rates (SFRs) 
derived from FIR emission
alone are suspect for quasars, as reprocessed quasar emission can, in 
principle, contribute to the FIR flux. In the 5/6 cases where
PAHs are detected, however, they show typical
ratios of PAH/FIR luminosities over a 
wide range of $L_{\rm FIR}$ (0.02-0.06 in total PAH integrated over our 
template, corresponding to 0.0014-0.0042 in the 6.2$\mu$m 
feature, within the range for star-forming galaxies and ULIRGs seen by 
Peeters, Spoon \& Tielens [2004] of $\sim 0.005-0.1$). 
This indicates 
that the bulk of the FIR emission is indeed likely to be 
from star formation (see also Sajina et al., in preparation). 
A similar result has been found for type-1 quasars by 
Schweitzer et al.\ (2006). As Figure 2 shows, however, 
the [O{\sc ii}]3727 emission (from the spectra of Lacy et al.\ [2007], 
aperture corrected using the SDSS 
photometry) is significantly below that 
expected in an unreddened starburst in most cases, 
particularly for the objects with high FIR luminosities. 
This is consistent with the result of Ho (2005),
who found low [O{\sc ii}]3727 emission from type-1 quasar hosts compared to 
their expected star formation rates based on measurements of molecular gas. 
However, the ratio of the 
[Ne{\sc ii}]12.8$\mu$m to the FIR luminosity
is similar for all our objects. 
[Ne{\sc ii}] is of similar ionization to [O{\sc ii}] and
is also therefore dominated by emission from the starburst. 
This suggests that dust
reddening is to blame for the lack of [O{\sc ii}]3727 emission, and is also
consistent with the
host galaxy colors and morphologies. As can be seen in Figure 1, the 
host galaxies of the objects with high $L_{\rm FIR}$ are typically disturbed, 
red objects, likely to be late stage merger remnants.
This suggests that star formation in a significant 
fraction of quasar host galaxies is heavily obscured in the optical.

\begin{figure}
\plotone{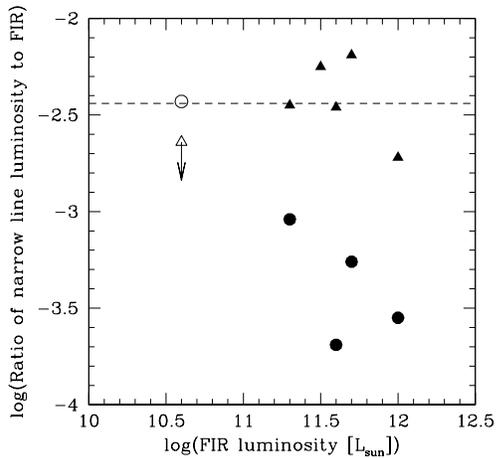}
\caption{The ratio of narrow line luminosity in [O{\sc ii}]3727 to 
FIR luminosity (circles) and in [Ne{\sc ii}]12.8$\mu$m to FIR
luminosity (triangles). Host galaxies 
dominated by a disk component are shown as filled symbols, 
SSTXFLS J172123.1+601214, which has an elliptical host, is shown with 
open symbols. The dashed line is the ratio
of [O{\sc ii}] to FIR luminosity expected in an unextincted starburst.}
\end{figure}

\begin{figure}
\plotone{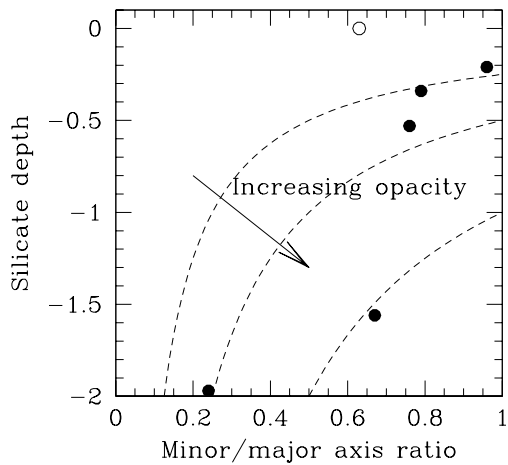}
\caption{Absorption depth in the 9.7$\mu$m silicate feature 
plotted against axial ratio. Symbols as for Figure 2. The dashed lines
represent the predictions of simple uniform disk models of differing 
opacity measured perpendicular to the disk.}
\end{figure}

Obscuration by dusty starforming regions also seems to affect the extinction
to the AGN. 
Figure 3 shows a plot of axial ratio of the host galaxy 
(measured using the {\em ellipse} task in {\sc iraf} on the longer 
wavelength image) versus the 
depth of the 9.7$\mu$m silicate feature, $S_{\rm sil}$,
as defined by Levenson et al.\ (2006). 
Although our sample is small, the five of our six host galaxies
with apparent or possible disk-like structures, and 1d-profiles
close to exponential disks (excluding morphological 
disturbances and bulge components), represented by the solid points, 
seem to show a trend in the sense that 
the more highly-extincted objects have more edge-on inclinations
(lower axial ratios). The one galaxy which does not appear disk-like,
SSTXFLS~J172123.1+601214 (open symbol), has an 
underlying $r^{1/4}$ law profile, and is probably an example of an elliptical 
host merging with a smaller gas-rich galaxy, and lacks the large
amount of star formation seen in the mergers of two gas-rich galaxies. In this
case, the obscuration is probably almost entirely due to the torus. 
For the disk-like hosts, a simple uniform dusty disk model appears to work for 
most, with a spread of about a factor of two in 
opacity for a line of sight perpendicular to the disk. One object
(SSTXLFS~J171106.8+590436), however, would have a high opacity in this model
even if viewed face-on, corresponding to $S_{\rm sil}\approx -1$. This
may be because the extinction is patchy, and a high inclination 
significantly increases the chance of the
line of sight passing through a high opacity region such as a dense molecular
cloud. Our two most obscured type-2 quasars (with $S_{\rm sil}=-1.6$ and $-2$)
have silicate absorption  as deep or deeper than the most obscured AGN in the 
sample of Shi et al.\ (2006). These
radio galaxies and Seyfert 2s have $S_{\rm sil}\geq -1.6$,
suggesting that extinction
can be very extreme in the radio quiet quasar population.

The implication of our result is that many type-2 quasars have at least a
component of the extinction towards the nucleus from an extended, star forming
disk on scales of kpc, in addition to, or instead of, 
the traditional obscuring torus, which interferometry suggests is only a few
pc in size (Jaffe et al.\ 2004).
Previous suggestions that the host galaxy can contribute significantly to the
AGN extinction in type-2 objects have
been made by  Martinez-Sansigre et al.\ (2006) and Rigby et 
al.\ (2006), based on optical emission line properties. 
Our results suggest that a significant fraction of the quasar population 
have their AGN obscured by starforming regions in their host galaxies. 
Thus studying obscured quasars is essential if
we are to understand the origin of the links between the evolution of 
quasars and their hosts.

\acknowledgments

We thank the referee for a helpful report.
ML and AOP were visiting astronomers at the IRTF, 
which is operated by the University of Hawaii under Cooperative Agreement 
no. NCC 5-538 with the National Aeronautics and Space Administration (NASA), 
Science Mission Directorate, Planetary Astronomy Program. 
This work is based on observations made with the Spitzer Space Telescope, 
which is operated by the Jet Propulsion Laboratory, California Institute of 
Technology under a contract with NASA. Support for this work was provided by 
NASA through an award issued by JPL/Caltech. This work is also based on 
observations made with the NASA/ESA Hubble Space Telescope, 
obtained at the Space Telescope Science Institute (STScI), operated by the 
Association of Universities for Research in Astronomy, Inc., under NASA 
contract NAS 5-26555, and are associated with program 10848, 
partly supported by a grant from STScI.

\begin{table*}
\caption{Host galaxy properties}
\scriptsize{
\begin{tabular}{lcccccccccc}
\hline
Object  & z&log$(L_{\rm AGN}$ & log($L_{\rm FIR}$ & log($L_{\rm PAH}$&log($L_{\rm [OII]}$&$S_{\rm sil}$ & Axial & SFR  & SFR\\
        &         &  $[L_{\odot}])$  &   $[L_{\odot}])$  & $[L_{\odot}])$  & $[L_{\odot}])$ &        &  ratio& (FIR)& ([O{\sc ii}])\dag \\ 
        &         &            &          & &&&&($M_{\odot}$yr$^{-1}$)&($M_{\odot}$yr$^{-1}$)\\\hline
SSTXFLS J172123.1+601214&0.325&11.4&10.6&$<$9.4&8.17& -   &0.63$\pm$0.02&3.3&3.9\\
SSTXLFS J171147.4+585839&0.800&12.1&11.7&  10.5&8.44&-0.21& 0.96$\pm$0.02  &56&7.2\\
SSTXFLS J171324.2+585549&0.609&11.8&11.3&   9.8&8.26&-0.34& 0.79$\pm$0.01  &16&4.8\\
SSTXFLS J171831.5+595317&0.700&11.9&12.0&  10.3&8.45&-0.53& 0.76$\pm$0.05  &87&7.4\\
SSTXFLS J171106.8+590436&0.462&11.8&11.5&  10.1&7.49&-1.56& 0.67$\pm$0.01  &33&0.8\\
SSTXFLS J172458.3+591545&0.494&11.6&11.6&  10.4&7.91&-1.97& 0.24$\pm$0.02  &37&2.1\\\hline
\end{tabular}

\noindent
$\dag$ from Kewley et al.\ (2004), assuming solar metallicity.
}
\end{table*}

\end{document}